\documentclass [prl,twocolumn,superscriptaddress,longbibliography]{revtex4-1}

\usepackage{graphicx}
\usepackage{subfigure}
\usepackage{amsmath}
\usepackage{amsthm}
\usepackage{amssymb}
\usepackage{hyperref}
\usepackage{xcolor}
\usepackage{textcomp}
\usepackage[normalem]{ulem}

\hypersetup{colorlinks=true, linkcolor=black, citecolor=black, urlcolor=blue}

\newcommand{\ket}[1]{\left| #1 \right>} % for Dirac bras
\newcommand{\bra}[1]{\left< #1 \right|} % for Dirac kets
\begin{document}

%=========================================================================
\title{Emergent Quasicrystalline Symmetry in Light-Induced Quantum Phase Transitions}
 
\author{Farokh Mivehvar}
\email[Corresponding author: ]{farokh.mivehvar@uibk.ac.at}
\affiliation{Institut f\"ur Theoretische Physik, Universit{\"a}t Innsbruck, A-6020~Innsbruck, Austria}
\author{Helmut Ritsch}
\affiliation{Institut f\"ur Theoretische Physik, Universit{\"a}t Innsbruck, A-6020~Innsbruck, Austria}
\author{Francesco Piazza}
\affiliation{Max-Planck-Institut f\"{u}r Physik komplexer Systeme, D-01187 Dresden, Germany}

\begin{abstract}
The discovery of quasicrystals with crystallographically forbidden rotational symmetries has changed
the notion of the ordering in materials, yet little is known about the dynamical emergence of such exotic forms of order. 
Here we theoretically study a nonequilibrium cavity-QED setup realizing 
a zero-temperature quantum phase transition from a homogeneous Bose-Einstein condensate 
to a quasicrystalline phase via collective superradiant light scattering. 
Across the superradiant phase transition, collective light scattering creates
a dynamical, quasicrystalline optical potential for the atoms.
Remarkably, the quasicrystalline potential is ``emergent'' as its eight-fold rotational symmetry
is not present in the Hamiltonian of the system, rather appears solely in the low-energy states.
For sufficiently strong two-body contact interactions between atoms, a quasicrystalline order is stabilized in the system,
while for weakly interacting atoms the condensate is localized in one or few of the deepest 
minima of the quasicrystalline potential.

\end{abstract}

\maketitle

%=========================================================================
\emph{Introduction.}---Quasicrystals are quasiordered (or orientationally ordered) materials with no exact translational symmetry, rather with crystallographically forbidden rotational symmetries~\cite{steinhardt_1984}. They possess rotational symmetries, such as five-, seven-, eight-fold rotational symmetries, as discovered from their diffraction patterns first by Shechtman \textit{et al.}\ in 1984~\cite{Shechtman1984Metallic}. Therefore, they are not periodic and do not belong to any of the crystallographic space groups. Interestingly, quasicrystals, related to aperiodic tilings, can be considered as projections of higher dimensional periodic lattices~\cite{deBruijn1981AlgebraicI, deBruijn1981AlgebraicII, Kramer1984On}. Despite extensive theoretical and experimental research since their discovery~\cite{Goldman1993Quasicrystals}, there are still many fundamental open questions concerning the formation and nature of quasicrystals. For instance, it is still not completely clear whether quasicrystals are only entropy-stabilized high-temperature states or can also be thermodynamically stable at low temperatures~\cite{Steurer2018Quasicrystals}. In particular, the conditions and nature of quasicrystal growth are under debate with a lack of a generally accepted model. 

%--------Figure------------ 
\begin{figure}[!b]
\centering
\includegraphics [width=0.49\textwidth]{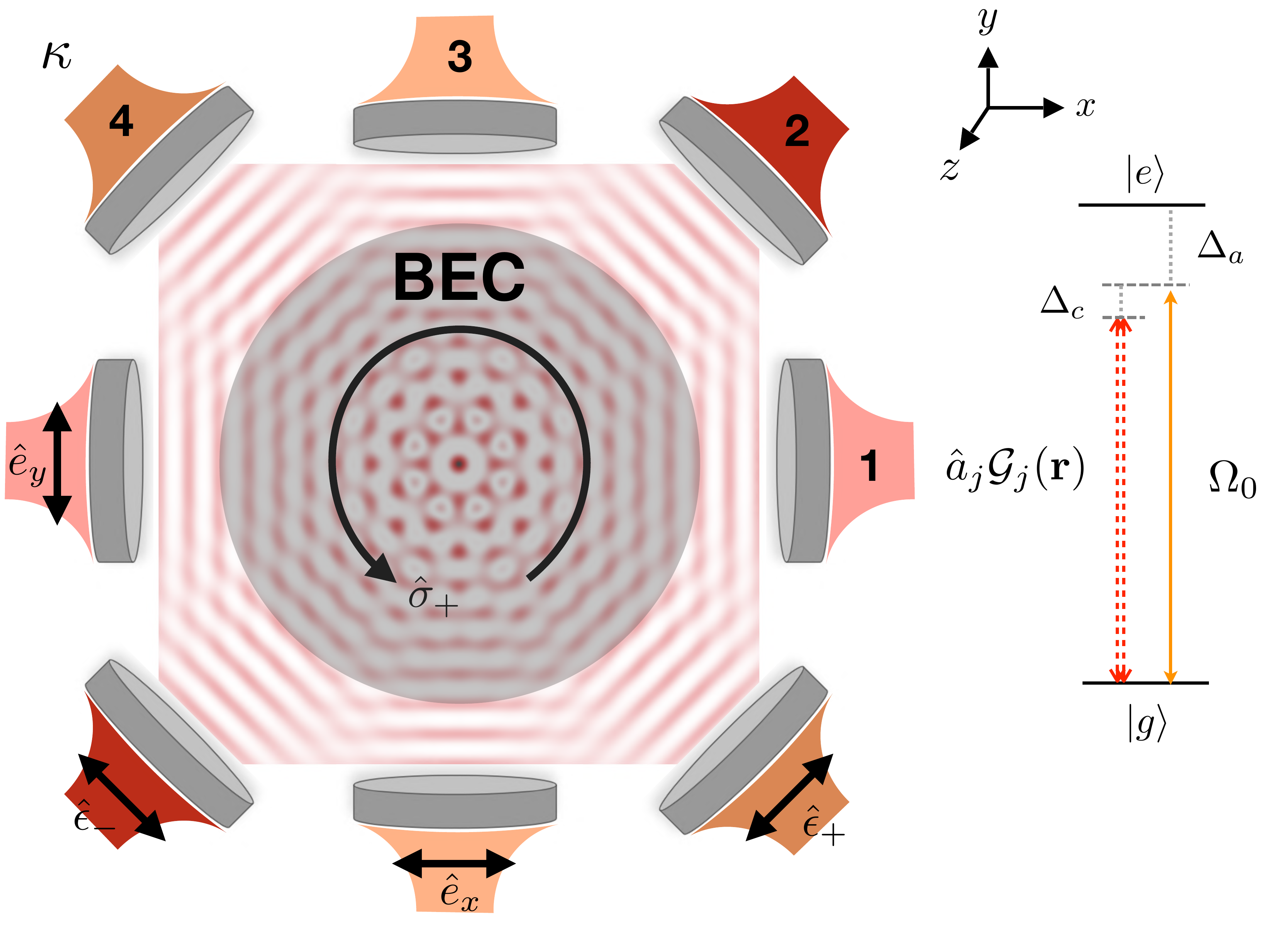}
\caption{Schematic view of a quasi-two dimensional driven BEC inside four identical crossed linear cavities,
which make 45\textdegree angle with one another.
The driving laser with the right circular polarization $\hat{\sigma}_+$ 
propagates along the $z$ direction and it is not shown explicitly in the figure.  
The inset depicts the internal atom-photon couplings.} 
\label{fig:qc-scheme}
\end{figure}

Ultracold atoms trapped in laser-created tailored optical potentials have proven to be a versatile platform for simulating and exploring exotic solid-state models in a controllable manner~\cite{Gross2017Quantum}. In this context, the recent loading of a Bose-Einstein condensate (BEC) into a specially designed quasicrystal optical potential led to the observation of a diffraction pattern with a forbidden eight-fold rotational symmetry and has opened a new perspective on studying quasicrystals~\cite{Viebahn2018Matter}.  This now allows for a detailed experimental investigation of the nature of quasicrystals and other theoretically predicted phenomena such as the interplay between quasiordering and localization~\cite{Sanchez-Palencia2005BEC-qc, Lye2007Effect, Cetoli2013Loss, Jagannathan2013an-eightfold, Bordia2017Probing, Spurrier2018Semiclassical, Hou2018Superfluid-Quasicrystal}. Unlike natural quasicrystals, however, the quasicrystallization in these systems is \textit{not} dynamically emergent and spontaneous~\cite{Goldstein1999Emergence, Corning2002re-emergence}, rather induced by the underlying optical potentials imposed on the atoms by prescribed external lasers. Recently, a few schemes based on spin-orbit-coupled BECs were proposed to realize quantum phase transitions to quasicrystals~\cite{Gopalakrishnan2013Quantum,Hou2018Superfluid}, where quasicrystalline rotational symmetries appear in these systems due to the interplay between the spin-orbit coupling and two-body interactions --- i.e., two competing length scales. This competition \cite{lifshitz_1997,denton_1998,engel_2007,barkan_2011,Rottler_2012,archer_2013,lifshitz_2014,Dotera2014Mosaic} has been identified to be also the relevant mechanism for the quasicrystallization in soft matter~\cite{jamie_2004,yushu_2007,Fischer1810,osamu_2012,framk_2012}.

In this Letter we propose an alternative, novel scenario for the spontaneous formation of a quasicrystal based on a dynamical optical potential for ultracold atoms inside an optical cavity setup~\cite{Ritsch2013Cold, Domokos2002Collective, Baumann2010Dicke, Schmidt2014Dynamical, Klinder2015Observation, Lonard2017Supersolid, Lonard2017Monitoring, Morales2018Coupling, Landini2018Formation, Kroeze2018Spinor, Guo2019Sign, Vaidya2018Tunable, Naik2018Bose, Schuster2018Pinning, Cosme2018Dynamical, Georges2018Light, Kroeze2019Dynamical, Habibian2013Bose, Ballantine2017Meissner, Lode2017Fragmented, Mivehvar2018Driven, Mivehvar2019Cavity,Chiacchio2019Dissipation}, where a quasicrystalline symmetry \textit{emerges} in the low-energy sector across a superradiant phase transition. The setup consists of four identical linear cavities arranged in a plane with a common center such that they make a 45\textdegree angle with one another. A  Bose gas is tightly confined in the direction perpendicular to the plane at the intersection of these \textit{initially empty} four cavities, and is strongly coupled to a single mode of each cavity. The BEC is also driven by a spatially uniform pump laser propagating perpendicular to the cavity-BEC plane as depicted in Fig.~\ref{fig:qc-scheme}. 

\begin{figure}[t!]
\centering
\includegraphics [width=0.49\textwidth]{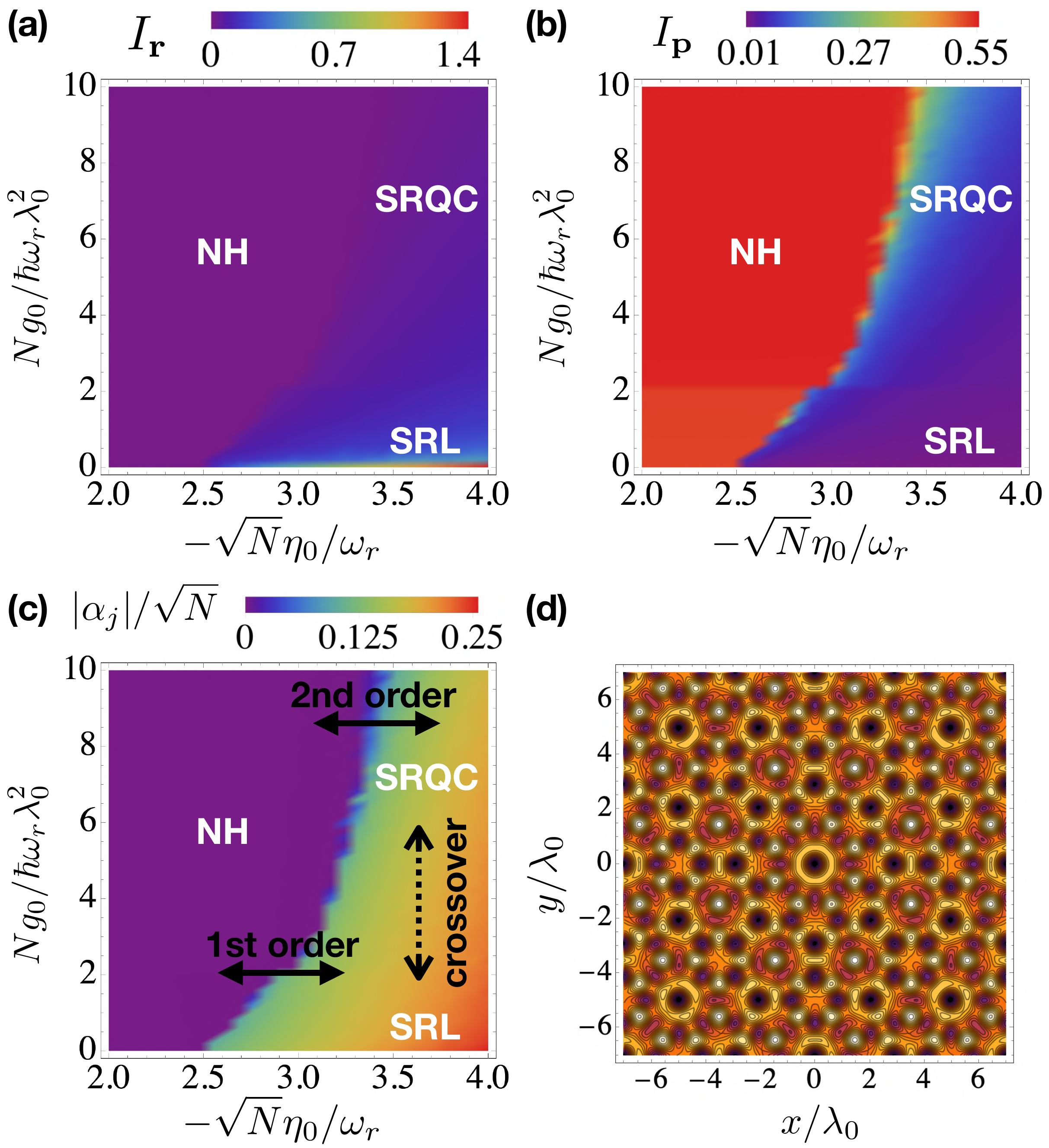}
\caption{The phase digram of the system in the $\{Ng_0/\hbar\omega_r\lambda_0^2,\sqrt{N}\eta_0/\omega_r\}$ parameter plane.
The system exhibits three distinct phases: normal homogenous (NH), superradiant localized (SRL), 
and superradiant quasicrystal (SRQC) states. The inverse participation ratios $\{I_{\mathbf r},I_{\mathbf p}\}$ 
as well as the field amplitudes $|\alpha_j|$ display non-analytical behavior on the onset of the superradiant phase transition as shown in (a)-(c).
The order and the nature of the transitions are indicated in (c).
A typical emergent superradiant quasicrystalline optical potential with the eight-fold rotational symmetry $C_8$
located at the origin is presented in (d).
The parameters are set to $(\Delta_c,\kappa,N\mathcal{G}_0^2/\Delta_a)=(-10,10,-1)\omega_r$.} 
\label{fig:phase_diags}
\end{figure}

At low pump-laser strengths, the system is in the normal homogenous (NH) state, where the condensate is uniform and the photon scattering from the pump into the cavities is strongly suppressed. Beyond a critical laser strength, atoms collectively scatter photons from the pump into the cavities as atomic density fluctuations are amplified due to the field backaction. Consequently, the pump and built-up cavity fields interfere, leading to the formation of a nonperiodic, dynamic superradiant quasicrystalline potential for the atoms, which possesses an \textit{emergent} eight-fold rotational symmetry $C_8$; see Figs.~\ref{fig:phase_diags}(c) and~(d). Although the Hamiltonian of the system does not possess this eight-fold rotational symmetry, across the superradiant phase transition this symmetry emerges in the low-lying energy sector thanks to specific cavity-field \textit{amplitudes} and \textit{phases} chosen spontaneously by the system. The center of the quasicrystal (i.e., the location of the $C_8$ rotational axis) is fixed via a process of spontaneous breaking of four approximate discrete $\mathbf{Z}_2$ symmetries. The mechanism of emerging global $C_8$ symmetry here is reminiscent of emergent global~\cite{ZAMOLODCHIKOV1989INTEGRALS, Coldea2010Quantum, Chen2015Quantum, Sagi2016Emergent, Lang_2017,Wu2018Emergent, Chiacchio2018Emergence, Guo2019Emergent} and local gauge~\cite{Baskaran1988Gauge, Senthil2004Deconfined, tHooft2007Emergent, Barcel2016From, Witten2018Symmetry} symmetries found in some other models in the proximity of certain quantum phase transitions and/or in some quantum phases.

\begin{figure}[t!]
\centering
\includegraphics [width=0.49\textwidth]{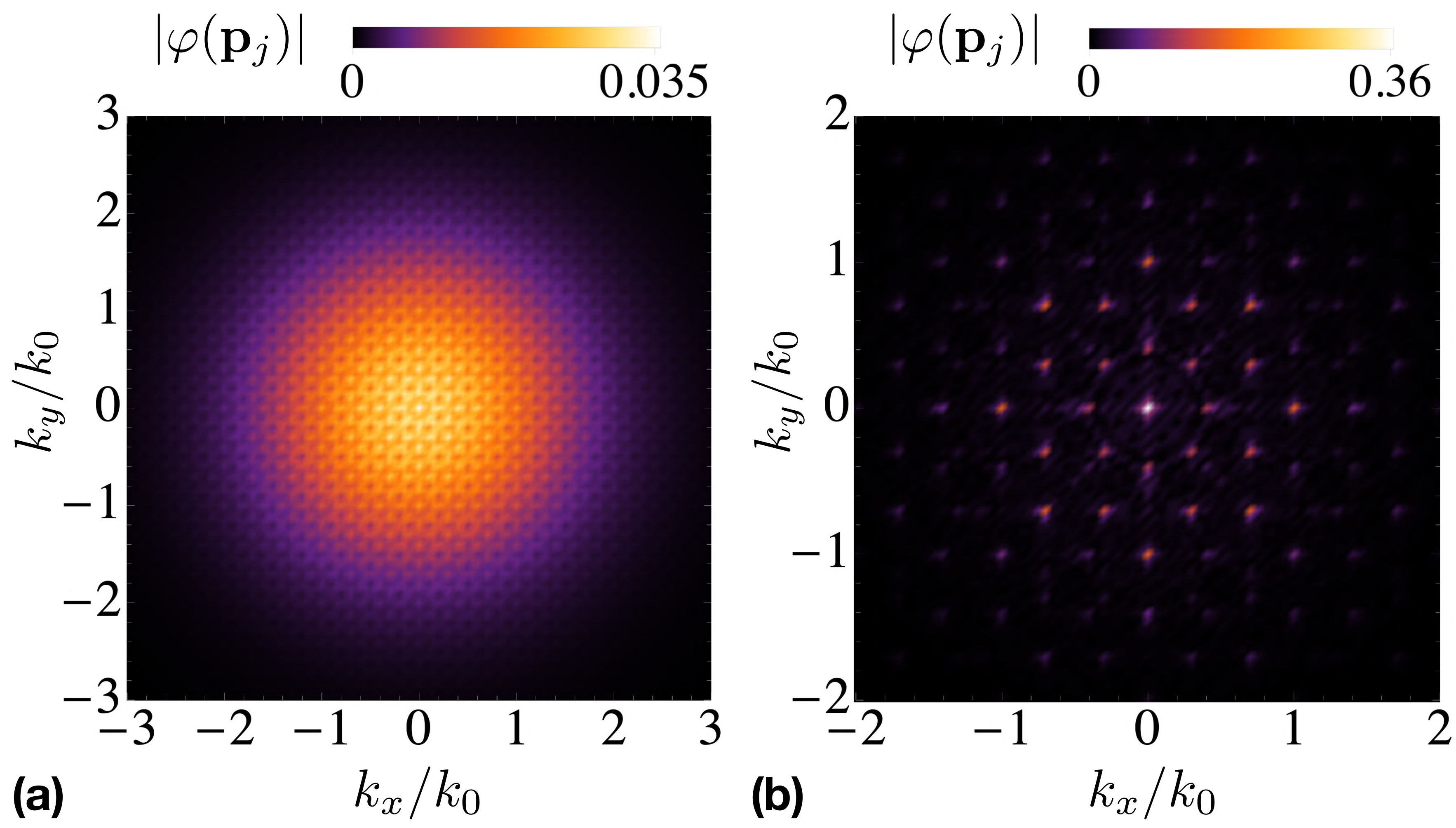}
\caption{Typical atomic momentum distribution in the SRL (a) and SRQC (b) phases.
The momentum distribution in the SRQC state exhibits a clear eight-fold rotational symmetry,
with occupied fractional momentum states. In the SRL phase, it is, however, a Gaussian with
remnants of the eight-fold rotational symmetry on top of it.
The parameters are set to $\sqrt{N}\eta_0/\omega_r=4$ and
$Ng_0/\hbar\omega_r\lambda_0^2=0$ (a) and $5$ (b), with the rest being the same as Fig.~\ref{fig:phase_diags}.
} 
\label{fig:mom-dist}
\end{figure}

In the superradiant phase, the atoms in turn self-order in this emergent quasicrystalline potential. For sufficiently strong two-body repulsive contact interactions between the atoms, a superradiant quasicrystalline (SRQC) order is stabilized in the system, where momentum components of the self-ordered  BEC wavefunction exhibit an eight-fold rotational symmetry as shown in Fig.~\ref{fig:mom-dist}(b). On the contrary, for weakly interacting atoms the condensate is localized in one or few of the deepest minima of the quasicrystalline potential. Correspondingly, the many occupied momentum components of the condensate wavefunction show a Gaussian distribution. We thus refer to this state as the superradiant localized (SRL) phase.

%=========================================================================
\emph{Model.}---Consider ultracold bosonic atoms trapped in a quasi-two-dimensional ``circular'' box potential $V_{\rm box}(\mathbf{r})$ in the $x$-$y$ plane and off-resonantly driven in the $z$ direction by a right circularly polarized $\hat\sigma_+$ pump laser with Rabi frequency $\Omega_0\propto\bra{e}\sigma_+\cdot\mathbf{d}\ket{g}$ and wave-number $k_0=2\pi/\lambda_0$. The atomic internal states $\{\ket{g},\ket{e}\}$ satisfy the selection rule $m_e-m_g=1$. Furthermore, the atomic transition $\ket{g}\leftrightarrow\ket{e}$ is also off-resonantly coupled to four \textit{initially empty}, in-plane polarized, quantized electromagnetic modes, each belonging to one linear cavity. The atom-cavity  couplings are given by $\mathcal{G}_j(\mathbf{r})=e^{i\theta_j}\mathcal{G}_{0j}\cos(\mathbf{k}_{j}\cdot\mathbf{r})$, where $\mathbf{k}_1=k_0\hat{e}_x$, $\mathbf{k}_3=k_0\hat{e}_y$, and $\mathbf{k}_{2,4}=k_0(\hat{e}_x\pm\hat{e}_y)/\sqrt{2}$, and $\mathcal{G}_{0j}$ are the maximum atom-cavity couplings per photon. Here, $\hat{e}_{x}$ ($\hat{e}_{y}$) is the unit vector along the $x$ ($y$) direction and $i^2=-1$.  The wave-number of the cavity modes have been assumed to be equal to the wave-number of the laser, $|\mathbf{k}_j|=k_0$, as the cavity frequencies $\omega_c\equiv\omega_{c1}=\cdots=\omega_{c4}$ are taken to be close to resonant with laser frequency $\omega_0=ck_0$. The phase factors $\theta_1=\pi/2$, $\theta_2=5\pi/4$, $\theta_3=\pi$, and $\theta_4=3\pi/4$ arise due to the projection of the in-plane linear polarizations $\hat{e}_{x,y}$ and $\hat{\epsilon}_{\pm}=(\hat{e}_x\pm\hat{e}_y)/\sqrt{2}$ of the cavity fields onto the right circular polarization $\hat\sigma_+$ as detailed in the Supplemental Material~\cite{SM-qc2019}. The system is depicted schematically in Fig.~\ref{fig:qc-scheme}.

Assuming that the atomic detuning $\Delta_a\equiv\omega_0-\omega_a<0$, with $\omega_a$ being the atomic transition frequency between the ground state $\ket{g}$ and the electronic excited state $\ket{e}$, is very large, the atomic excited state can be adiabatically eliminated. This yields an effective Hamiltonian $\hat{H}_{\rm eff}=\int \hat{\psi}^\dag(\mathbf{r}) \hat{\mathcal H}_{0,\rm eff}\hat{\psi}(\mathbf{r})d\mathbf{r}+\hat{H}_{\rm int}-\hbar\Delta_c\sum_j\hat{a}^\dag_j\hat{a}_j$ for the system, with the effective single-particle Hamiltonian density in the rotating-frame of the laser,
\begin{align} \label{eq:q-1-H}
\hat{\mathcal H}_{0,\rm eff}=-\frac{\hbar^2}{2M}\nabla^2+V_{\rm box}(\mathbf{r})
+\frac{\hbar}{\Delta_a}\Big |\Omega_0+\sum_{j=1}^4\hat{a}_j\mathcal{G}_j(\mathbf{r})\Big |^2,
\end{align}
and the two-body interaction Hamiltonian,
\begin{align}
\hat{H}_{\rm int}=g_0\int \hat{\psi}^\dag(\mathbf{r})\hat{\psi}^\dag(\mathbf{r})\hat{\psi}(\mathbf{r})\hat{\psi}(\mathbf{r})d\mathbf{r}.
\end{align}
Here $\hat{\psi}(\mathbf{r})$ and $\hat{a}_j$ are the atomic and photonic bosonic field operators, respectively, $\Delta_c\equiv\omega_0-\omega_c$ is the cavity detunings with respect to the lasers, and $g_0$ is the strength of the two-body contact interactions proportional to the s-wave scattering length. 

%=========================================================================
\emph{Mean-field approach.}---We consider the thermodynamic limit, where the quantum fluctuations are negligible in two dimensions and the mean-field approach is justified~\cite{Piazza2013Bose}: $\hat\psi\to\psi\equiv\langle\hat\psi\rangle$ and $\hat{a}_j\to\alpha_j=|\alpha_j|e^{i\gamma_j}\equiv\langle\hat{a}_j\rangle$. Hence we solve the mean-field Gross-Pitaevskii equation 
\begin{align} \label{eq:GP-eq}
\left(\langle\hat{\mathcal H}_{0,\rm eff}\rangle+g_0n(\mathbf{r})\right)\psi(\mathbf{r})=\mu\psi(\mathbf{r}),
\end{align}  
along with the self-consistent solution of the steady-state cavity-field amplitudes $i\hbar\langle\partial_t\hat{a}_j\rangle=\langle[\hat{a}_j,\hat{H}_{\rm eff}]\rangle-i\hbar\kappa\langle\hat{a}_j\rangle=0$, 
\begin{align} \label{eq:ss-alphas}
-\delta_{cj}\alpha_j+\sum_{\ell\neq j}c_{j\ell}\alpha_\ell+\eta_j=0; \qquad j,\ell=1,2,3,4.
\end{align}
Here, $\langle\hat{\mathcal H}_{0,\rm eff}\rangle$ is the mean-field single-particle Hamiltonian density corresponding to Eq.~\eqref{eq:q-1-H} with $\hat{a}_j\to\alpha_j$, $n(\mathbf{r})=|\psi(\mathbf{r})|^2$ is the local atomic density, $\mu$ is the chemical potential, $\kappa$ is the decay rate of the cavity fields due to photon losses through the cavity mirrors, and we have introduced the following symbols for the shorthand,
\begin{align} \label{eq:shorthands}
\delta_{cj}&=\Delta_c+i\kappa-\frac{1}{\Delta_a}\int |\mathcal{G}_j(\mathbf{r})|^2n(\mathbf{r})d\mathbf{r},\nonumber\\
c_{j\ell}&=\frac{1}{\Delta_a}\int \mathcal{G}_j^*(\mathbf{r})\mathcal{G}_\ell(\mathbf{r})n(\mathbf{r})d\mathbf{r},\nonumber\\
\eta_j&=\frac{\Omega_0}{\Delta_a}\int \mathcal{G}_j^*(\mathbf{r})n(\mathbf{r})d\mathbf{r}.
\end{align}
This approach neglects the heating induced by cavity losses, which is well justified as the corresponding rate is suppressed with the inverse system's size~\cite{piazza_2014}. Without loss of generality and for the sake of simplicity, in the following we set $\mathcal{G}_{0j}$ and $\Omega_{0}$ to be real and only focus on the most interesting case of symmetric coupling to all cavities, $\mathcal{G}_0\equiv\mathcal{G}_{01}=\cdots=\mathcal{G}_{04}$.

Since the system is not translationally invariant, we solve Eqs.~\eqref{eq:GP-eq} and \eqref{eq:ss-alphas} in a square box of size $L_x\times L_y=L^2=20\lambda_0\times20\lambda_0$ centered at the origin $\mathbf{r}=0$ with open boundary conditions and a circular box potential of the form $V_{\rm box}(\mathbf{r})=0$ for $|\mathbf{r}|<L/2$, otherwise, $V_{\rm box}(\mathbf{r})\to\infty$~\footnote{One can use instead an azimuthally symmetric harmonic trap in an experiment. The discrete $\mathbf{Z}_2$ symmetries discussed in the text are then absent in the system and the center of the emergent quasicrystal potential is fixed externally at the center of the trap.}.
 In order to characterize the BEC density and momentum distributions, we exploit the inverse participation ratios~\cite{Thouless1974,Roux2008qpBH}, 
 \begin{subequations}
\begin{align} \label{eq:q_drop}
I_{\mathbf r}&=\int |\psi(\mathbf{r})|^4d\mathbf{r}/\Big(\int |\psi(\mathbf{r})|^2d\mathbf{r}\Big)^2,
\end{align}
\begin{align} \label{eq:q_drop}
I_{\mathbf p}&=\sum_{\mathbf{p}_j} |\varphi(\mathbf{p}_j)|^4/\Big(\sum_{\mathbf{p}_j} |\varphi(\mathbf{p})|^2\Big)^2,
\end{align}
\end{subequations}
where $\varphi(\mathbf{p})=\int e^{i\mathbf{p}\cdot\mathbf{r}} \psi(\mathbf{r})d\mathbf{r}/L^2$ is the Fourier transform of the condensate wavefunction $\psi(\mathbf{r})$. $I_{\mathbf r}$ and $I_{\mathbf p}$ quantify how localized the atomic distributions are in position and momentum spaces, respectively. For instance, for a uniform density distribution $I_{\mathbf r}$ approaches zero in the thermodynamic limit as $1/L^2$, while $I_{\mathbf p}$ approaches one.

%=========================================================================
\emph{Phase diagram and emergent symmetries.}---Figure~\ref{fig:phase_diags} shows the phase diagram of the system in the $\{Ng_0/\hbar\omega_r\lambda_0^2,\sqrt{N}\eta_0/\omega_r\}$ parameter plane, where $N$ is the number of the atoms, $\omega_r\equiv\hbar k_0^2/2M$ is the recoil frequency, and $\eta_0\equiv\mathcal{G}_0\Omega_0/\Delta_a$ is the effective pump strength~\footnote{The somewhat rugged boundaries are partially due to finite-size effects as discussed in Supplemental Material. Furthermore, parameter grids in the phase diagrams were somewhat coarse as the numerical simulations were quite time consuming due to the lack of periodic boundary conditions.}. The inverse participation ratios in position $I_{\mathbf r}$ and momentum $I_{\mathbf p}$ space are illustrated in Figs.~\ref{fig:phase_diags}(a) and \ref{fig:phase_diags}(b), respectively. The corresponding rescaled steady-state field amplitudes $|\alpha_j|/\sqrt{N}$ are presented in Fig.~\ref{fig:phase_diags}(c).

%--------Figure------------ 

Below the laser-strength threshold $\eta_0^c(g_0)$ the system is in the normal homogenous (NH) state, where there is no photon in any of the cavities and the atoms are uniformly (saving for the boundary) distributed over the box potential $V_{\rm box}$. Therefore, the position-space (momentum-space) participation ratio $I_{\mathbf r}$ ($I_{\mathbf p}$) assumes its smallest (largest) value in this phase. 

By increasing the pump-laser strength $\eta_0$ above the threshold $\eta_0^c(g_0)$, the uniform atomic distribution starts to become unstable. Fluctuations in the atomic density result in constructive photon scattering from the pump laser into the cavity modes. The interference of the pump and built-up cavity fields creates an emergent optical potential, favoring density modulations which in turn further enhance collective photon scattering into the cavity modes. This starts a runaway process towards a superradiant phase where $|\alpha_j|>0$ for all $j$. Both participation ratios change at the onset of the superradiant phase transition. In particular, the momentum-space participation ratio $I_{\mathbf p}$ displays a sharp drop, signaling the occupation of many momentum states. In the superradiant phase $\alpha_j=|\alpha_j|e^{i\gamma_j}\neq0$, not only are the absolute values of all the field amplitudes equal $|\alpha|\equiv|\alpha_1|=\cdots=|\alpha_4|\neq0$, but also the phase of each field amplitude is locked at $\gamma_j=\gamma_0-\theta_j$ or $\gamma_0+\pi-\theta_j$, where $\gamma_0$ is a common phase introduced due to the nonzero cavity-field decay rate $\kappa\neq0$. 

The single-particle Hamiltonian density $\hat{\mathcal H}_{0,\rm eff}$, Eq.~\eqref{eq:q-1-H}, possesses a symmetry which we denote it by $\tilde{C}_8$: an eight-fold rotation $C_8$ around the $z$ axis with the transformation $x\to x'=(x+y)/\sqrt{2}$ and $y\to y'=(y-x)/\sqrt{2}$ followed by the field transformation $\hat{a}_1\to\hat{a}_2e^{-i(\theta_1-\theta_2)}$, $\hat{a}_2\to\hat{a}_3e^{-i(\theta_2-\theta_3)}$, $\hat{a}_3\to\hat{a}_4e^{-i(\theta_3-\theta_4)}$, and $\hat{a}_4\to\hat{a}_1e^{-i(\theta_4-\theta_1)}$. Note that $\hat{\mathcal H}_{0,\rm eff}$ is \textit{not} invariant under the sole action of the $C_8$ rotation. Nonetheless, an eight-fold rotational symmetry $C_8$ \textit{emerges} in the low-energy sector of the superradiant phase due to the above-mentioned amplitude and phase locking of the cavity fields: $|\alpha|\equiv|\alpha_1|=\cdots=|\alpha_4|\neq0$, and  $\gamma_j+\theta_j=0$ or $\pi$ ($\gamma_0$ is an immaterial overall common phase shift discarded here for the sake of simplicity); see Supplemental Material for more details~\cite{SM-qc2019}. The two possible choices of phase, $-\theta_j$ or $\pi-\theta_j$, for each field amplitude correspond to a $\mathbf{Z}_2$ symmetry for each cavity. The phases $\gamma_j$ of the field amplitudes determine the center of the quasicrystal (i.e., the location of the $C_8$ rotational axis). Figure~\ref{fig:phase_diags}(d) shows an example for the case $\gamma_j=-\theta_j$ for $j=1,\cdots,4$. Therefore, at the onset of the superradiant phase transition, the center of the emergent quasicrystal is fixed through the spontaneous breaking of an approximate $\otimes_{j=1}^4\mathbf{Z}_2$ symmetry as explained in detail in the following.

For the case $\gamma_j=-\theta_j$ ($\gamma_j=\pi-\theta_j$) for all the modes $j=1,\cdots,4$, all fields have positive (negative) antinode at the origin $\mathbf{r}=0$. Therefore, the $C_8$ rotational axis is located at the origin, defining the center of the quasicrystal. Whereas, e.g., for the case that $\gamma_1=\pi-\theta_1$ and $\gamma_j=-\theta_j$ for the rest (i.e., $j=2,3,4$), the rotational axis is shifted along the $x$ axis and its location $x_o$ has to satisfy both $x_o=(m+1/2)\lambda_0$ and $x_o=\sqrt{2}l\lambda_0$, with $m$ and $l$ being two integers. However, these two conditions are not exactly consistent with each other for any two finite integers $m$ and $l$, as the first equation yields a rational number while the second one an irrational number. One might, therefore, claim that the two conditions may coincide at infinity, implying that the center of the quasicrystal is located at $x_o\to\pm\infty$.

%--------Figure------------ 

The self-ordered condensate density in the superradiant phase depends on the interplay between the emergent quasicrystal optical potential and the two-body repulsive contact interaction $g_0$. Since the optical potential is not periodic, it can act as a disordered potential for the BEC, favoring Anderson-type localization~\cite{Roati2008Anderson, Billy2008Direct}. On the other hand, this is counteracted by the two-body repulsive interactions. This competition determines the self-ordered BEC density. 

In the weakly interacting regime, the quasicrystal potential dominates over the two-body interactions and the condensate localizes in one or few of deepest minima of the quasicrystal potential. In this localized phase the atoms scatter more photons from the pump laser into the cavity modes due to constructive interference: the more atoms are condensed in the same potential minimum, the more photons are scattered with the same phase. This superradiant localized (SRL) regime is the lower right corner in Figs.~\ref{fig:phase_diags}(a)-(c). The inverse participation ratio in the position space $I_{\mathbf r}$ attains its largest values in this phase. A typical momentum distribution of the self-ordered SRL state is shown in Fig.~\ref{fig:mom-dist}(a); strong localization here leads to a Gaussian momentum distribution with remnants of the underlying eight-fold rotational symmetry superimposed on it.

In the strongly interacting regime on the other hand, it is energetically more favorable for the atoms to occupy different global and local minima of the quasicrystal optical potential. For sufficiently strong two-body contact interactions, a quasicrystalline density order is stabilized in the system as apparent from the momentum distribution of self-ordered states, shown in Fig.~\ref{fig:mom-dist}(b) for a typical state in this regime. Note the fractal distribution, a characteristic of quasicrystalline order indicating that the two-dimensional momentum space cannot be spanned by only two reciprocal primitive vectors. Here the two-dimensional momentum space is spanned by four incommensurate cavity wavevectors $\{\mathbf{k}_1,\cdots,\mathbf{k}_4\}$, resulting in dense and self-similar momentum diffraction peaks~\cite{Viebahn2018Matter}. Furthermore, regardless of the center of the quasicrystal, all the sixteen $\otimes_{j=1}^4\mathbf{Z}_2$ symmetry-broken states exhibit an identical momentum distribution, saving for small finite-size effects~\cite{SM-qc2019}. Therefore, we refer to this phase as the superradiant quasicrystalline (SRQC) state.

One can identify the field amplitude $|\alpha|$ (recall that $|\alpha|\equiv|\alpha_1|=\cdots=|\alpha_4|$) as an order parameter. It is zero in the NH phase and acquires nonzero values in the SRL and SRQC phases. The order parameter $|\alpha|$ exhibits a discontinuous jump in the transition from the NH phase to the SRL state, signaling a first order phase transition. The transition from the NH state to the SRQC phase is also first order in the weakly interacting regime, but becomes second order in the strongly interacting regime. On the other hand, we have only a crossover between the SRL and SRQC phases, as the order parameter $|\alpha|$  changes smoothly in the transition. For clarity, various cuts from the phase diagrams are presented in Supplemental Material for different transitions~\cite{SM-qc2019}.

%=========================================================================
\emph{Conclusions.}---We proposed and studied a novel cavity-QED setup where a quantum phase transition from a uniform BEC to a quasicrystalline state with an emergent eight-fold rotational symmetry can be realized and monitored non-destructively via the amplitudes $|\alpha_1|=\cdots=|\alpha_4|\neq0$ and phases $\gamma_j=\gamma_0-\theta_j$ or $\gamma_0+\pi-\theta_j$ of the cavity-output fields. The proposed setup is a realistic generalization of the state of the art in the experiment, namely the two-crossed cavity setup~\cite{Lonard2017Supersolid,Lonard2017Monitoring,Morales2018Coupling} and the bow-tie cavities~\cite{Naik2018Bose}. Analogous setups can allow the study of quasicrystals with other emergent rotational symmetries such as five- and seven-fold rotational symmetries.
Our proposed model differs from all previously proposed schemes including those based on spin-orbit coupling~\cite{Gopalakrishnan2013Quantum, Hou2018Superfluid}: in the previous proposals the quasicrystalline states have a \textit{broken} symmetry with respect to corresponding single-particle Hamiltonians while here the superradiant quasicrystalline state has an \textit{emergent} symmetry. Furthermore, the quantum phase transition to the quasicrystalline and localized states can be attributed to the interplay between two-body contact interaction and cavity-mediated long-range interactions. Therefore, our work demonstrates that cavity QED offers a novel non-demolishing platform~\cite{Mekhov2007Probing, Mekhov2007Cavity, Gietka2019Supersolid, Torggler2019A} for exploring the spontaneous formation and stabilization mechanisms of a quasicrystal at very low temperatures. It opens a new avenue for realizing in state-of-the-art quantum-gas--cavity systems quantum phase transitions which are accompanied by an emergent ($C_8$ rotational) symmetry. Our considered system may also support exotic vestigial orders in some parameter regimes~\cite{Gopalakrishnan2017Intertwined}.

%=========================================================================
\begin{acknowledgments}
We acknowledge inspiring discussions with Julian L\'eonard.
F.~M.\ is grateful to Stefan Ostermann for fruitful discussions regarding numerics.
F.~M.\ is supported by the Lise-Meitner Fellowship
M2438-NBL of the Austrian Science Fund (FWF), and the
International Joint Project No.\ I3964-N27 of the FWF and
the National Agency for Research (ANR) of France.
%H.~R.\ acknowledges support from \textcolor{red}{XXXX}.
\end{acknowledgments}

%=========================================================================
%\bibliographystyle{aip}
\bibliography{qc}

%=========================================================================
\newpage
\widetext
\setcounter{equation}{0}
\setcounter{figure}{0}
\renewcommand{\theequation}{S\arabic{equation}}
\renewcommand{\thefigure}{S\arabic{figure}}

%=========================================================================
\section{Supplemental Material}

Here we present the details of the derivation of the superradiant optical potential for the atoms, represented by the last term in Eq.~(1) in the manuscript. We then discuss the symmetries of the system. The effect of the finite-size system in the symmetry-broken states are investigated numerically. In addition, we present some cuts from the phase diagrams to see more clearly the order and the nature of the phase transitions. Finally we show some typical atomic density distributions.

%=========================================================================
\section{The Effective Optical Potential} 

The total electric field is the sum of the pump and cavity fields,
\begin{align} \label{eq:E-original}
\hat{\mathbf E}(\mathbf{r})=E_{0p}\hat\sigma_++\sum_{j=1}^4\mathcal{E}_{0j}(\hat{a}_j+\hat{a}_j^\dag)\cos(\mathbf{k}_j\cdot\mathbf{r}+\phi_j)\hat{\epsilon}_j,
\end{align}
where the electric field per photon $\mathcal{E}_{0j}=\mathcal{E}_0=(\hbar\omega_c/\epsilon_0 V)^{1/2}$ is the same for all the cavities, $\mathbf{r}=(x,y)$, $\mathbf{k}_1=k_0\hat{e}_x$, $\mathbf{k}_3=k_0\hat{e}_y$, and $\mathbf{k}_{2,4}=k_0(\hat{e}_x\pm\hat{e}_y)/\sqrt{2}$, with $\hat{e}_{x}$ ($\hat{e}_{y}$) being the unit vector along the $x$ ($y$) direction. We consider in-plane linear polarizations for the cavity fields, so that $\hat{\epsilon}_1=\hat{e}_y$, $\hat{\epsilon}_3=\hat{e}_x$, $\hat{\epsilon}_{2,4}=\hat{\epsilon}_\mp=(\hat{e}_x\mp\hat{e}_y)/\sqrt{2}$. These linear polarizations can be expressed in the basis of circular polarizations $\hat\sigma_\pm=\mp(\hat{e}_{x}\pm i\hat{e}_{y})/\sqrt{2}$ as,
\begin{align}
\hat{e}_x&=\frac{1}{\sqrt{2}}(-\hat\sigma_++\hat\sigma_-),\nonumber\\
\hat{e}_y&=\frac{i}{\sqrt{2}}(\hat\sigma_++\hat\sigma_-),\nonumber\\
\hat{\epsilon}_+&=\frac{1}{\sqrt{2}}(e^{3i\pi/4}\hat\sigma_++e^{i\pi/4}\hat\sigma_-),\nonumber\\
\hat{\epsilon}_-&=\frac{1}{\sqrt{2}}(e^{5i\pi/4}\hat\sigma_++e^{7i\pi/4}\hat\sigma_-).
\end{align}
Therefore, the electric field \eqref{eq:E-original} can be recast as $\hat{\mathbf E}(\mathbf{r})=\hat{E}_+(\mathbf{r})\hat\sigma_++\hat{E}_-(\mathbf{r})\hat\sigma_-$, where
\begin{align} \label{eq:E_+}
\hat{E}_+(\mathbf{r})&=E_{0p}+\frac{\mathcal{E}_0}{\sqrt{2}}\Big[
e^{i\pi/2}(\hat{a}_1+\hat{a}_1^\dag)\cos(\mathbf{k}_1\cdot\mathbf{r})
+e^{5i\pi/4}(\hat{a}_2+\hat{a}_2^\dag)\cos(\mathbf{k}_2\cdot\mathbf{r})\nonumber\\
&\hskip 2.1cm+e^{i\pi}(\hat{a}_3+\hat{a}_3^\dag)\cos(\mathbf{k}_3\cdot\mathbf{r})
+e^{3i\pi/4}(\hat{a}_4+\hat{a}_4^\dag)\cos(\mathbf{k}_4\cdot\mathbf{r})
\Big],\nonumber\\
\hat{E}_-(\mathbf{r})&=\frac{\mathcal{E}_0}{\sqrt{2}}\Big[
e^{i\pi/2}(\hat{a}_1+\hat{a}_1^\dag)\cos(\mathbf{k}_1\cdot\mathbf{r})
+e^{7i\pi/4}(\hat{a}_2+\hat{a}_2^\dag)\cos(\mathbf{k}_2\cdot\mathbf{r})\nonumber\\
&\hskip 1.1cm+(\hat{a}_3+\hat{a}_3^\dag)\cos(\mathbf{k}_3\cdot\mathbf{r})
+e^{i\pi/4}(\hat{a}_4+\hat{a}_4^\dag)\cos(\mathbf{k}_4\cdot\mathbf{r})
\Big].
\end{align}
We have assumed that the cavity fields all have an antinode at the origin $\mathbf{r}=(0,0)$ by setting $\phi_j=0$ for all $j$.

Let us now consider two atomic internal states $\{\ket{g},\ket{e}\}$, such that their magnetic quantum numbers satisfy $m_e-m_g=1$. The Rabi frequency is then given by 
\begin{align} \label{eq:Rabi-freq}
\hat{\Omega}(\mathbf{r})&=\bra{e}\hat{\mathbf E}(\mathbf{r})\cdot\mathbf{d}\ket{g}
=\hat{E}_+(\mathbf{r})\bra{e}\hat\sigma_+\cdot\mathbf{d}\ket{g}\nonumber\\
&\simeq\Omega_{0}+\mathcal{G}_0\Big[
e^{i\pi/2}\hat{a}_1\cos(\mathbf{k}_1\cdot\mathbf{r})
+e^{5i\pi/4}\hat{a}_2\cos(\mathbf{k}_2\cdot\mathbf{r})
+e^{i\pi}\hat{a}_3\cos(\mathbf{k}_3\cdot\mathbf{r})
+e^{3i\pi/4}\hat{a}_4\cos(\mathbf{k}_4\cdot\mathbf{r})
\Big],
\end{align}
where the rotating-wave approximation has been used in the last equality. For a large atomic detuning $\Delta_a$, this leads to a quantized optical potential (i.e., position-dependent Stark shift) for the atoms,
\begin{align} \label{eqSM:Vqc}
\hat{V}(\mathbf{r})=\frac{\hbar}{\Delta_a}|\hat{\Omega}(\mathbf{r})|^2=
\frac{\hbar}{\Delta_a}\Big|\Omega_0+\sum_{j=1}^4\hat{a}_j\mathcal{G}_j(\mathbf{r})\Big|^2.
\end{align}

%=========================================================================
\section{Symmetry Considerations}

The effective single-particle Hamiltonian density of the system in the rotating-frame of the laser is given by,
\begin{align} \label{eqSM:1-H}
\hat{\mathcal H}_{0,\rm eff}=-\frac{\hbar^2}{2M}\nabla^2+V_{\rm box}(\mathbf{r})
+\hat{V}(\mathbf{r}).
\end{align}
Let us consider a $C_8$ rotational axis along the $z$ axis located at the origin $\mathbf{r}=(x,y)=(0,0)$ with the transformation $x\to x'=(x+y)/\sqrt{2}$ and $y\to y'=(y-x)/\sqrt{2}$. The optical potential operator, Eq.~\eqref{eqSM:Vqc}, is then transformed under the $C_8$ rotational symmetry as
\begin{align} \label{eqSM:Vqc-C8}
\hat{V}(\mathbf{r})\to \hat{V}'(\mathbf{r})=
\frac{\hbar}{\Delta_a}\Big|\Omega_0+\hat{a}_1e^{i(\theta_1-\theta_2)}\mathcal{G}_2(\mathbf{r})+\hat{a}_2e^{i(\theta_2-\theta_3)}\mathcal{G}_3(\mathbf{r})+\hat{a}_3e^{i(\theta_3-\theta_4)}\mathcal{G}_4(\mathbf{r})+\hat{a}_4e^{i(\theta_4-\theta_1)}\mathcal{G}_1(\mathbf{r})\Big|^2
\neq\hat{V}(\mathbf{r}).
\end{align} 
Hence, the single-particle Hamiltonian density $\hat{\mathcal H}_{0,\rm eff}$ is \textit{not} invariant under the $C_8$ rotational symmetry. Instead, $\hat{\mathcal H}_{0,\rm eff}$ is invariant under a larger symmetry operator which we denote it by $\tilde{C}_8$: the $C_8$ rotational symmetry followed by the field transformation,
\begin{align}
\hat{a}_1\to\hat{a}_2e^{-i(\theta_1-\theta_2)},\nonumber\\
\hat{a}_2\to\hat{a}_3e^{-i(\theta_2-\theta_3)},\nonumber\\
\hat{a}_3\to\hat{a}_4e^{-i(\theta_3-\theta_4)},\nonumber\\
\hat{a}_4\to\hat{a}_1e^{-i(\theta_4-\theta_1)}.
\end{align}
 
For a generic state with field amplitudes $\{\alpha_1,\cdots,\alpha_4\}$ where $\alpha_1\neq\alpha_2\neq\alpha_3\neq\alpha_4$, the superradiant optical potential 
\begin{align}
V(\mathbf{r})=\langle \hat{V}(\mathbf{r}) \rangle= 
\frac{\hbar}{\Delta_a}\Big|\Omega_0+\sum_{j=1}^4\alpha_j\mathcal{G}_j(\mathbf{r})\Big|^2,
\end{align}
also does not possess a $C_8$ rotational symmetry. However, the superradiant optical potential $V(\mathbf{r})$ becomes invariant under the $C_8$ rotational symmetry located at the origin \textit{if and only if} the field amplitudes take the specific values $\alpha_j=|\alpha|e^{-i\theta_j}$ for $j=1,\cdots,4$ (i.e., specific absolute values \textit{and} phases; here for the sake of simplicity of the notation we discard the overall common phase $\gamma_0$ introduced due to the nonzero cavity-field decay rate):
\begin{align} \label{eqSM:VMFqc-C8}
V(\mathbf{r})\to V'(\mathbf{r})=
\frac{\hbar}{\Delta_a}\Big|\Omega_0&+|\alpha|e^{-i\theta_1}e^{i(\theta_1-\theta_2)}\mathcal{G}_2(\mathbf{r})+|\alpha|e^{-i\theta_2}e^{i(\theta_2-\theta_3)}\mathcal{G}_3(\mathbf{r})\nonumber\\
&+|\alpha|e^{-i\theta_3}e^{i(\theta_3-\theta_4)}\mathcal{G}_4(\mathbf{r})+|\alpha|e^{-i\theta_4}e^{i(\theta_4-\theta_1)}\mathcal{G}_1(\mathbf{r})\Big|^2
=V(\mathbf{r}).
\end{align} 
In general, one can show for states with field amplitudes $\alpha_j=|\alpha|e^{i\gamma_j}$ where $\gamma_j=-\theta_j$ or $\gamma_j=\pi-\theta_j$ that the superradiant optical potential $V(\mathbf{r})$ possesses a $C_8$ rotational symmetry, but located at different positions in each case (note that $\gamma_j=-\theta_j$ for all $j=1,\cdots,4$ corresponds to the case discussed above where the  $C_8$ rotational symmetry is located at the origin). These sixteen states constitute the $\otimes_{j=1}^4\mathbf{Z}_2$ symmetry-broken, low-lying energy states of the system.

Simply stated, arranging four identical linear cavities with a 45\textdegree\ angle with one another in a plane does not imply the formation of a quasicrystalline potential with an eight-fold rotational symmetry $C_8$ in the superradiant phase \textit{a priori}. That is, for a generic solution of the mean-field Heisenberg field equations, with the corresponding steady-state field equations shown in Eq.~(4) in the manuscript, with field amplitudes $\{\alpha_1,\cdots,\alpha_4\}$ where $\alpha_1\neq\alpha_2\neq\alpha_3\neq\alpha_4$, the superradiant optical potential does not possess a $C_8$ rotational symmetry. The quasicrystalline potential with a $C_8$ rotational symmetry \textit{emerges} in the superradiant phase only for specific states with field amplitudes $\alpha_j=|\alpha|e^{i\gamma_j}$ where $\gamma_j=-\theta_j$ or $\pi-\theta_j$, which are the $\otimes_{j=1}^4\mathbf{Z}_2$ symmetry-broken, low-energy states of the system.

%=========================================================================
\section{Finite-Size Effects}

\begin{figure*}[t!]
 \centering
 \includegraphics [width=0.99\textwidth]{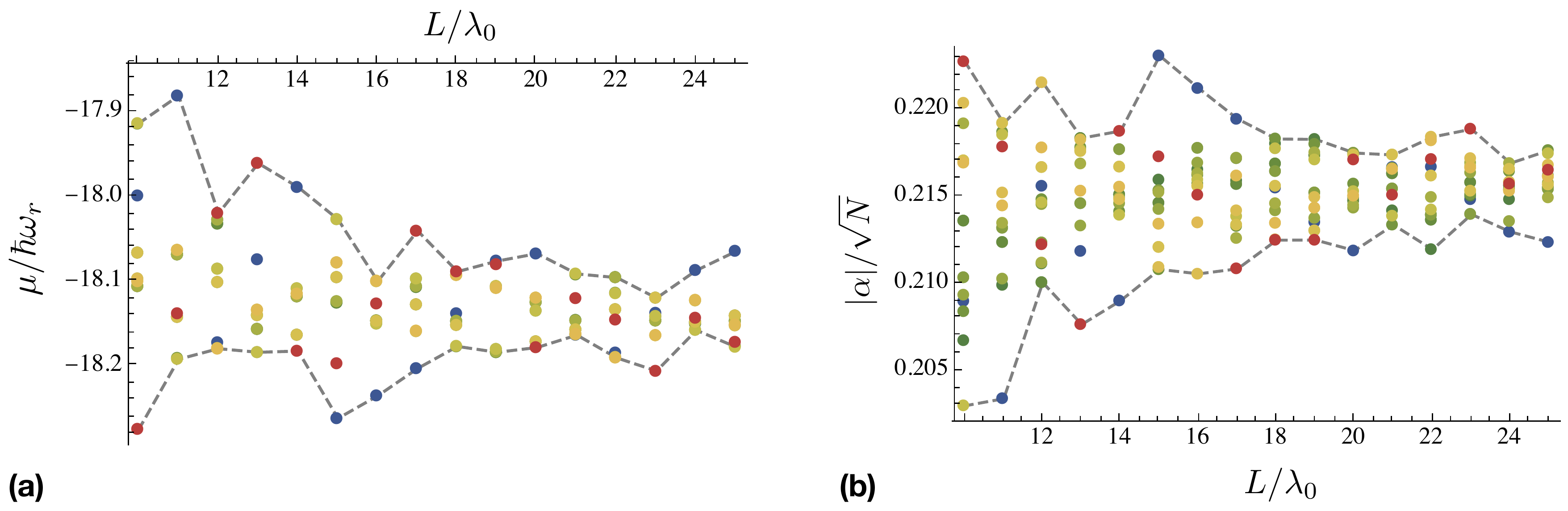}
 \caption{The finite-size effect.
(a) The chemical potential $\mu$ of the sixteen $\otimes_{j=1}^4\mathbf{Z}_2$ 
 symmetry-broken states as a function of the system size $L$ while keeping $Ng_0/L^2$ constant. 
 The energy difference between sixteen symmetry-broken states is a finite-size effect as
 it decreases by increasing the system size. 
 (b) The corresponding self-determined (rescaled) field amplitude $|\alpha|/\sqrt{N}$  
 as a function of the system size $L$.
 The red (blue) data points correspond to 
 the symmetry-broken state with $\gamma_j=\gamma_0-\theta_j$ ($\gamma_j=\gamma_0+\pi-\theta_j$)
 for all the modes $j=1,\cdots,4$. The other colors represent the other symmetry-broken states.
 The dashed lines are guides to the eye.
 The parameters are set to $\sqrt{N}\eta_0/\omega_r=4$ 
 and $Ng_0/\hbar\omega_r\lambda_0^2=1$ for the system size $L^2=(10\lambda_0)^2$, 
 with the rest being the same as Fig.~2 in the main text.
 } 
 \label{fig:mu_L}
 \end{figure*}

Let us now consider effects of the finite-size system in the sixteen $\otimes_{j=1}^4\mathbf{Z}_2$ symmetry-broken states. In an infinite system, all sixteen symmetry-broken states would have the same energy. However, in a finite system this is not necessarily true, as for each symmetry-broken state the center of the quasicrystal is located in a different place, and in fact it might not be even within the finite-size system. Therefore, the amplitude of scattered photons from the pump into the cavities is different for each symmetry-broken state, leading to lifting of the degeneracy of the ground-state manifold. Figures~\ref{fig:mu_L}(a) and \ref{fig:mu_L}(b) show, respectively, the chemical potential $\mu$ of the sixteen $\otimes_{j=1}^4\mathbf{Z}_2$ symmetry-broken states and the corresponding field amplitudes $\alpha$ as a function of the system size $L$ while keeping $Ng_0/L^2$ constant. As the system size increases, the field amplitudes and the energies of all the states converge. 

In a finite system only one of the sixteen symmetry-broken states is the true stationary (ground) state and the others are only quasi-stationary (i.e., metastable) states. Our imaginary-time-propagation simulations suggest the lifetime of these quasi-stationary states are infinitely long and they are almost stationary. That is, when the imaginary-time propagation converges randomly to any of the symmetry-broken states, it remains in that solution for the whole simulation time. This implies that energy barriers separating these states are very large and these states are metastable states of the system.

In obtaining the phase diagrams presented in Fig.~2 in the manuscript, numerical simulation has been implemented such that it can converge randomly to any of the sixteen $\otimes_{j=1}^4\mathbf{Z}_2$ symmetry-broken states in each point in the parameter space of the phase diagrams. Combined with the fact that the superradiant threshold could be slightly different for each symmetry-broken state, this has resulted in somewhat rugged phase boundaries. Another reason for the rugged phase boundaries is somewhat coarse parameter grids in the phase diagrams, chosen to optimize the numerical simulation time. The numerical simulations were quite expensive time-wise, due to the lack of periodic boundary conditions.

 %--------Figure------------ 

%=========================================================================
\section{Order and Nature of the Quantum Phase Transitions}

\begin{figure*}[t!]
 \centering
 \includegraphics [width=0.98\textwidth]{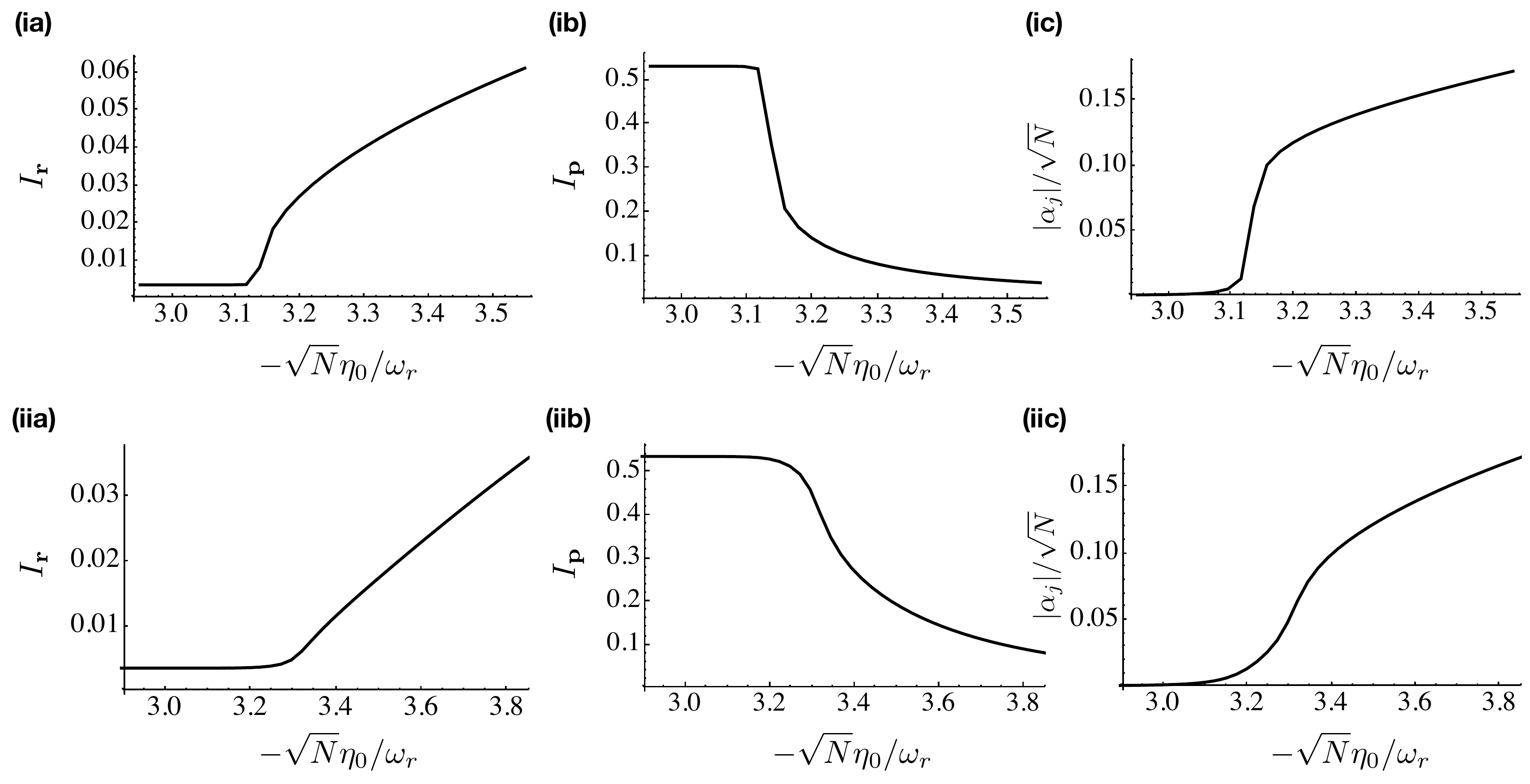}
 \caption{Horizontal cuts of the phase diagrams for $Ng/\hbar\omega_r\lambda_0^2=4$ (ia)-(ic) 
 and 9 (iia)-(iic).
 The other parameters are the same as Fig.~2 in the main text.
 } 
 \label{fig:PD_HC}
 \end{figure*}

\begin{figure*}[t!]
 \centering
 \includegraphics [width=0.98\textwidth]{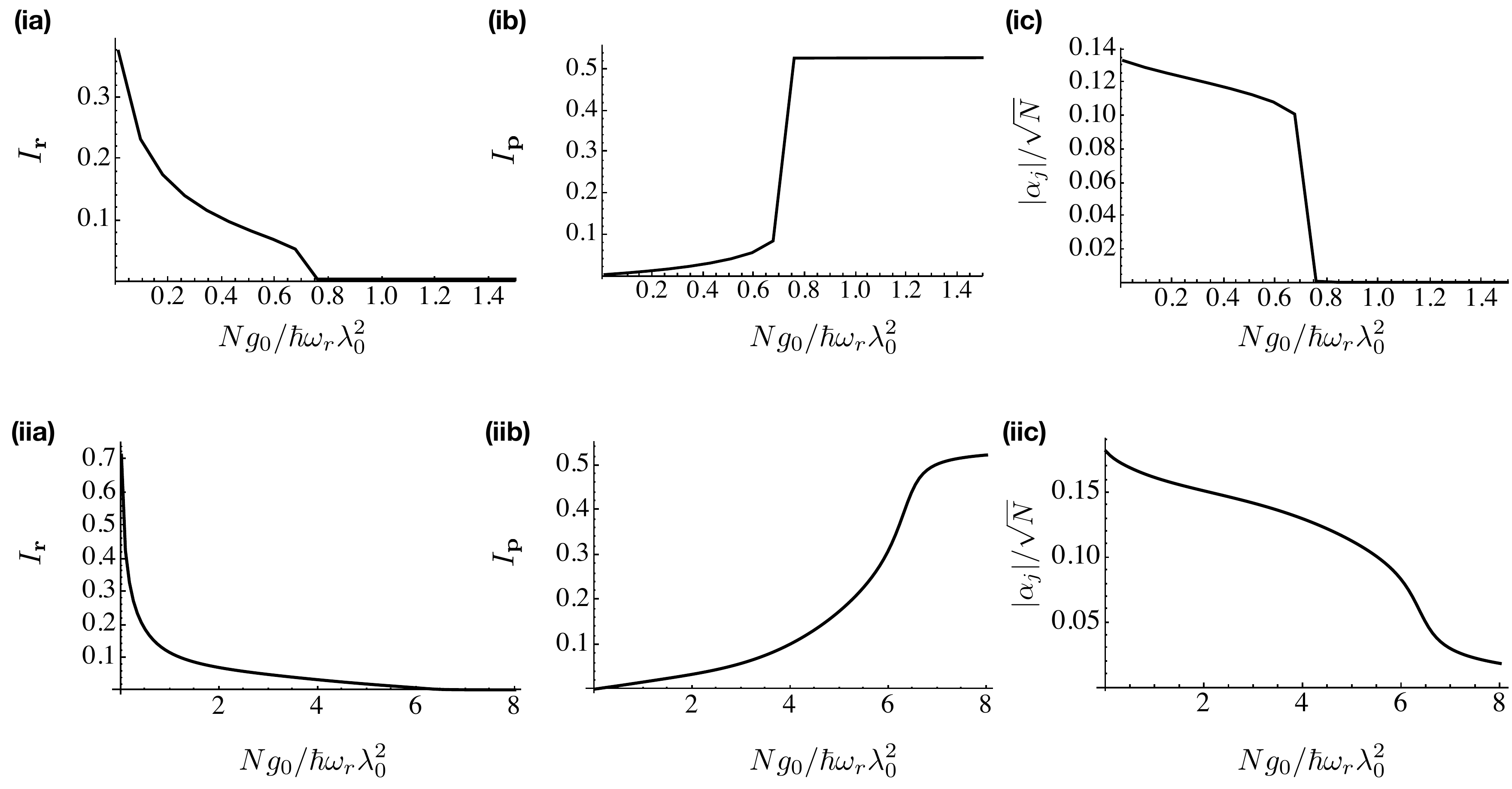}
 \caption{Vertical cuts of the phase diagrams for $\sqrt{N}\eta_0/\omega_r=2.7$ (ia)-(ic) 
 and $3.2$ (iia)-(iic).
 The other parameters are the same as Fig.~2 in the main text.
 } 
 \label{fig:PD_VC}
 \end{figure*}

In order to see the order and the nature of the quantum phase transitions more clearly, in Figs.~\ref{fig:PD_HC} and \ref{fig:PD_VC} we present, respectively, horizontal and vertical cuts from the phase diagrams of Fig.~2 in the main text. As can be seen from Fig.~\ref{fig:PD_HC}, in the weakly interacting regime the quantum phase transitions from the NH phase to the SRL and SRQC phases are first order as the order parameter $|\alpha|$ (recall that $|\alpha|\equiv|\alpha_1|=...=|\alpha_4|$) exhibits a discontinuous jump. On the other hand, in the strongly interacting regime the quantum phase transition from the NH state to the SRQC phase is second order. Note the lack of a sharp phase transition point in this case: it is due to the finite size of the system. The transition will become sharp in the thermodynamic limit. The transition from the SRL phase to the SRQC state is a crossover instead, since the order parameter $|\alpha|$ changes smoothly in the transition as can be seen from Fig.~\ref{fig:PD_VC}. For the sake of completeness, we have also illustrated the inverse participation ratios $\{I_{\mathbf r},I_{\mathbf p}\}$.

%--------Figure------------ 

 %--------Figure------------ 

%=========================================================================
\section{Density Distribution}

\begin{figure*}[t!]
 \centering
 \includegraphics [width=0.85\textwidth]{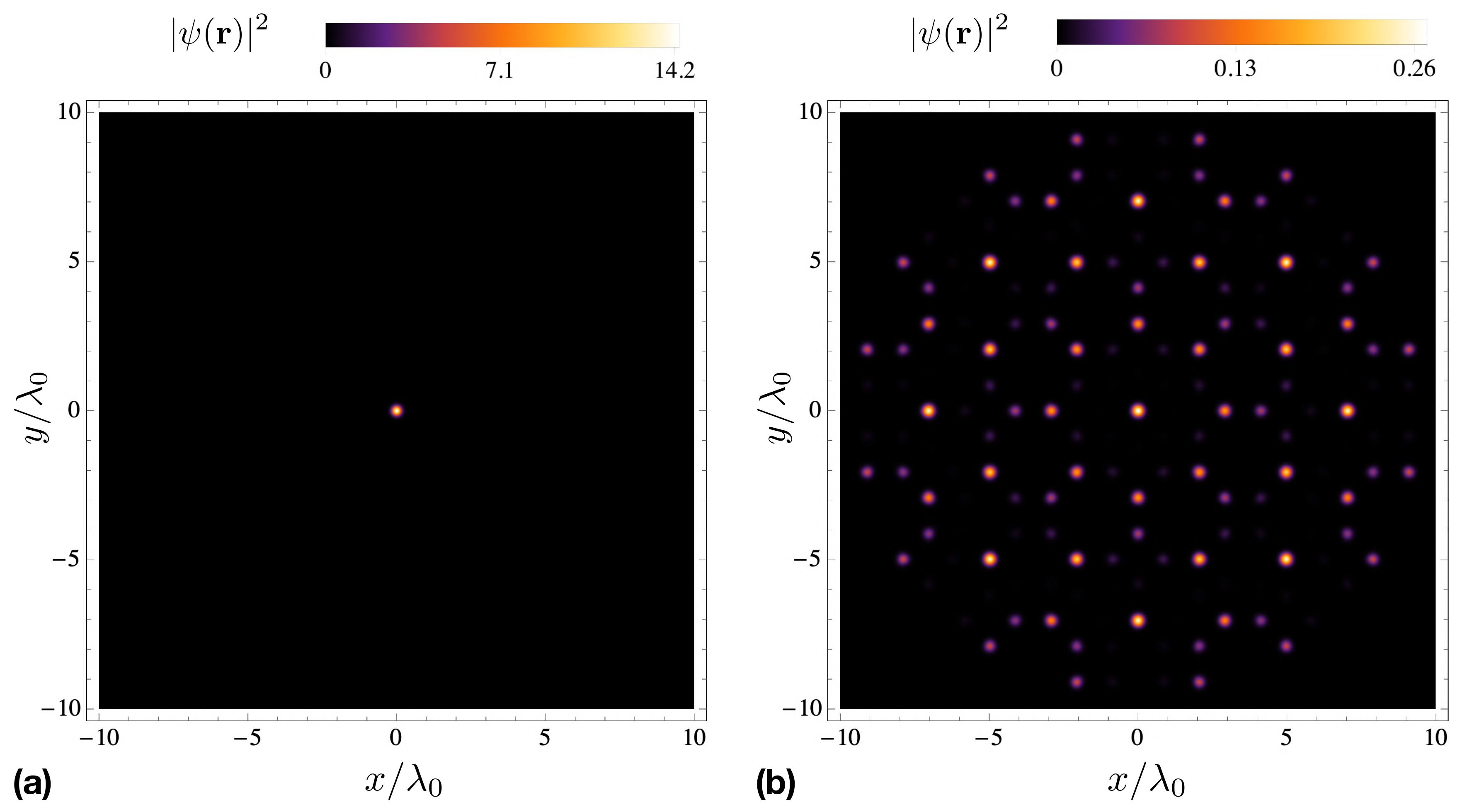}
 \caption{
Typical atomic density distribution in the SRL (a) and SRQC (b) phases.
The parameters are set to $\sqrt{N}\eta_0/\omega_r=4$ and
$Ng_0/\hbar\omega_r\lambda_0^2=0$ (a) and $5$ (b), corresponding
to the atomic momentum distributions shown in Fig.~3 in the manuscript.
The other parameters are the same as Fig.~2.
 } 
 \label{fig:den_dist}
 \end{figure*}

Typical atomic density distributions $|\psi(\mathbf{r})|^2$ in the SRL and SRQC phases are shown in Figs.~\ref{fig:den_dist}(a) and \ref{fig:den_dist}(b), respectively. In Fig.~\ref{fig:den_dist}(b), the center of the quasicrystal is located at the origin $\mathbf{r}=0$ --- corresponding to $\gamma_j=\gamma_0-\theta_j$ for all the modes --- exhibiting the eight-fold rotational symmetry manifestly. In the SRL state shown in Fig.~\ref{fig:den_dist}(a), almost all of the atoms condense into the global potential minimum in the center of the quasicrystal. 

 %--------Figure------------ 

%=========================================================================

\end{document}